\documentclass[]{llncs}

\usepackage{amsfonts}
\usepackage{color}
\usepackage{amsmath}

\usepackage{xargs}
\usepackage{ifthen}

\usepackage{longtable}
\usepackage{booktabs}
\usepackage{multirow}
\usepackage{rotating}

\usepackage{paralist}

\usepackage{multirow}
\usepackage[bookmarks=false]{hyperref}

\def\S#1{\mbox{{\sf \emph{#1}}}}
\def\K#1{\textbf{#1}}
\def\N{\\[.1ex]}                                       
\def\I{\hspace{1em}}

\newcommand{\myparagraph}[1]{\par\noindent\vspace{0.5mm}{\bf#1}}
\newcommand{\Id}[1]{\ensuremath{\mathit{#1}}}

\title{HordeSat: A Massively Parallel Portfolio SAT Solver}
\author{
	Tom\'{a}\v{s} Balyo, Peter Sanders, Carsten Sinz
		\thanks{This research was partially supported by DFG project SA 933/11-1}
}
\institute{
Karlsruhe Institute of Technology (KIT) \\
Karlsruhe, Germany
}

\begin{document}
\pagestyle{plain}

\maketitle

\begin{abstract}
A simple yet successful approach to parallel satisfiability (SAT) solving
is to run several different (a portfolio of) SAT solvers on the input problem 
at the same time until one solver finds a solution. The SAT solvers in
the portfolio can be instances of a single solver with different configuration
settings. Additionally the solvers can exchange information usually in the form of clauses.
In this paper we investigate whether this approach is applicable in the case
of massively parallel SAT solving. Our solver is intended to run on clusters with
thousands of processors, hence the name HordeSat.
HordeSat is a fully distributed portfolio-based SAT solver with a modular design
that allows it to use any SAT solver that implements a given interface.
HordeSat has a decentralized design and features hierarchical parallelism with
interleaved communication and search.
We experimentally evaluated it using all the benchmark problems from the
application tracks of the 2011 and 2014 International SAT Competitions.
The experiments demonstrate that HordeSat is scalable up to hundreds or even thousands of processors
achieving significant speedups especially for hard instances.
\end{abstract}


\section{Introduction}
\label{sec:introduction}
Boolean satisfiability (SAT) is one of the most important problems of theoretical
computer science with many practical applications in which SAT solvers are used
in the background as high performance reasoning engines. These applications include
automated planning and scheduling~\cite{kautz1992planning}, formal verification~\cite{aigs}, and
automated theorem proving~\cite{flanagan2003theorem}.
In the last decades the performance of state-of-the-art SAT solvers has increased
dramatically thanks to the invention of advanced heuristics~\cite{moskewicz2001chaff}, 
preprocessing and inprocessing
techniques~\cite{inprocessing} and data structures that allow efficient implementation
of search space pruning~\cite{moskewicz2001chaff}.

The next natural step in the development of SAT solvers was parallelization.
A very common approach to designing a parallel SAT solver is to run several instances of a sequential SAT
solver with different settings (or several different SAT solvers) on the same problem in parallel.
If any of the solvers succeeds in finding a solution all the solvers are terminated.
The solvers also exchange information mainly in the form of learned clauses.
This approach is referred to as portfolio-based parallel SAT solving and was first used
in the SAT solver ManySat~\cite{hamadi2008manysat}.
However, so far it was not clear whether this approach can scale to a large number of processors.

Another approach is to run several search procedures in parallel and ensure that they work on disjoint
regions of the search space. This explicit search space partitioning has been used mainly in solvers
designed to run on large parallel systems such as clusters or grids of computers \cite{chrabakh2003gridsat}.

In this paper we describe HordeSat -- a scalable portfolio-based
SAT solver and evaluate it experimentally. Using efficient yet thrifty clause exchange
and advanced diversification methods, we are able to keep the search spaces largely disjoint without
explicitly splitting search spaces.  
Another important feature of HordeSat is its modular design, which
allows it to be independent of any concrete search engines. HordeSat uses Sat
solvers as black boxes communicating with them via a minimalistic interface.

Experiments made using benchmarks from the application tracks of the 2011 and 2014 Sat
Competitions \cite{belov2014sat} show that HordeSat can outperform
state-of-the-art parallel SAT solvers on multiprocessor machines and is
scalable on computer clusters with thousands of processors. Indeed, we even
observe superlinear average speedup for difficult instances.


\section{Preliminaries}
\label{sec:preliminaries}

A \emph{Boolean variable} is a variable with two possible values \emph{True} and \emph{False}. 
By a~\emph{literal} of a Boolean variable $x$ we mean either $x$ or $\overline{x}$ (\emph{positive} 
or \emph{negative literal}). A~\emph{clause} is a disjunction (OR) of literals. 
A \emph{conjunctive normal form \textsc{(CNF)} formula} is a conjunction (AND) of clauses.
A clause can be also interpreted as a set of literals and a formula as a set of clauses.
A truth assignment $\phi$ of a formula $F$ assigns a truth value to its variables. The assignment $\phi$ satisfies a positive (negative) literal if it assigns the value True (False) to its variable and $\phi$ satisfies a clause if it satisfies any of its literals. Finally, $\phi$ satisfies a CNF formula if it satisfies all of its clauses. A formula $F$ is said to be satisfiable if there is a truth assignment $\phi$ that satisfies $F$. Such an assignment is called a \emph{satisfying assignment}. The satisfiability problem (\textsc{SAT}) is to find a satisfying assignment of a given CNF formula or determine that it is unsatisfiable. 

\subsubsection*{Conflict Driven Clause Learning.}
Most current complete state-of-the-art SAT solvers are based on the conflict-driven clause 
learning (CDCL) algorithm~\cite{marques1999grasp}. In this paper we will use CDCL solvers only as black boxes
and therefore we provide only a very coarse-grained description. For a detailed discussion of CDCL refer
to~\cite{biere2009conflict}.
In Figure~\ref{fig:cdcl} we give a pseudo-code of CDCL. The algorithm performs a depth-first search
of the space of partial truth assignments (\texttt{assignDecisionLiteral}, 
\texttt{backtrack} -- unassigns variables) interleaved with
search space pruning in the form of unit propagation (\texttt{doUnitPropagation}) and learning new clauses when the search reaches a conflict
state (\texttt{analyzeConflict}, \texttt{addLearnedClause}). If a conflict cannot be resolved by backtracking
then the formula is unsatisfiable. If all the variables are assigned and no conflict is detected
then the formula is satisfiable.

\begin{figure}[bt]
\centerline{\begin{minipage}{.7\linewidth}
{\centering \sf
\def\L#1{\raise .2ex\hbox{\tiny\tt #1}&}
\begin{tabular}{r@{\quad}l}
\L{} \S{CDCL} (CNF formula $F$) \N
\L{CDCL0} \I \K{while} not all variables assigned \K{do}\N
\L{CDCL1} \I \I assignDecisionLiteral \N
\L{CDCL2} \I \I doUnitPropagation \N
\L{CDCL3} \I \I \K{if} conflict detected \K{then} \N
\L{CDCL4} \I \I \I analyzeConflict \N
\L{CDCL5} \I \I \I addLearnedClause \N
\L{CDCL6} \I \I \I backtrack \K{or} \K{return} UNSAT \N
\L{CDCL7} \I \K{return} SAT \N
\end{tabular}}
\end{minipage}}
\caption{Pseudo-code of the conflict-driven clause learning (CDCL) algorithm.}
\label{fig:cdcl}
\end{figure}

\section{Related Work}
In this section we give a brief description of previous parallel SAT solving approaches. 
A much more detailed listing and description of existing parallel 
solvers can be found in recently published overview papers such as 
\cite{holldobler2011short,martins2012overview}.

\subsubsection*{Parallel CDCL -- Pure Portfolios.}
The simplest approach is to run CDCL several times on the same problem in parallel
with different parameter settings and exchanging learned clauses.
If there is no explicit search space partitioning then this approach is referred to as the pure
portfolio algorithm. 
The first parallel portfolio SAT solver was ManySat \cite{hamadi2008manysat}. The winner of the
latest (2014) Sat Competition's parallel track -- Plingeling \cite{biere13lingeling} is also of this kind.

The motivation behind the portfolio approach is that the performance of CDCL is heavily influenced
by a high number of different settings and parameters of the search
such as the heuristic used to select a decision literal.
Numerous heuristics can be used in this step~\cite{moskewicz2001chaff} 
but none of them dominates all the other heuristics
on each problem instance. Decision heuristics are only one of the many settings that
strongly influence the performance of CDCL solvers. All of these settings can be considered
when the diversification of the portfolio is performed. For an example see ManySat~\cite{hamadi2008manysat}.
Automatic configuration of SAT solvers in order to ensure that the solvers in a portfolio are 
diverse is also studied \cite{hydra}.

Exchanging learned clauses grants an additional boost of performance. It is an important
mechanism to reduce duplicate work, i.e., parallel searches working on the same part of the search
space. A clause learned from a conflict by one CDCL instance distributed to all the other CDCL
instances will prevent them from doing the same work again in the future.

The problem related to clause sharing is to decide how many and which clauses
should be exchanged. Exchanging all the learned clauses is infeasible especially
in the case of large-scale parallelism. A simple solution is to distribute all
the clauses that satisfy some conditions. The conditions are usually related to
the length of the clauses and/or their glue value~\cite{audemard2009predicting}.
An interesting technique called ``lazy clause exchange'' was introduced in a recent
paper~\cite{audemard2014lazy}. We leave the adaptation of this technique to future work however, since it
would make the design of our solver less modular.  Most of the existing pure
portfolio SAT solvers are designed to run on single multi-processor computers.
An exception is CL-SDSAT \cite{hyvarinen2014incorporating} which is designed for
solving very difficult instances on loosely connected grid middleware. It is not
clear and hard to quantify whether this approach can yield significant speedups
since the involved sequential computation times would be huge.

\subsubsection*{Parallel CDCL -- Partitioning The Search Space Explicitly.}
The classical approach to parallelizing SAT solving is to split the search space between
the search engines such that no overlap is possible. This is usually done by
starting each solver with a different fixed partial assignment. If a solver discovers that
its partial assignment cannot be extended into a solution it receives a new assignment.
Numerous techniques have presented how to manage the search space splitting
based on ideas such guiding paths \cite{chrabakh2003gridsat}, work stealing 
\cite{jurkowiak2005parallelization}, and generating
sufficiently many tasks \cite{gil2008pmsat}. Similarly to the portfolio approach the solvers
exchange clauses.

Most of the previous SAT solvers designed for computer clusters or grids 
use explicit search space partitioning.
Examples of such solvers are GridSAT \cite{chrabakh2003gridsat}, PM-SAT \cite{gil2008pmsat}, 
GradSat \cite{chrabakh2003gradsat},
C-sat \cite{ohmura2009csat}, ZetaSat \cite{blochinger2005zetasat} and SatCiety \cite{schulz2010parallel}.
Experimentally Comparing HordeSat with those solvers is problematic, since these solvers
are not easily available online or they are implemented for special environments using non-standard middleware.
Nevertheless we can get some conclusions based on looking at the
experimental sections of the related publications.

Older grid solvers such as GradSat~\cite{chrabakh2003gradsat}, PM-SAT~\cite{gil2008pmsat}
SatCiety~\cite{schulz2010parallel}, ZetaSat~\cite{blochinger2005zetasat} and C-sat~\cite{ohmura2009csat} 
are evaluated on only small clusters (up to 64 processors) using small sets of 
older benchmarks, which are easily solved
by current state-of-the-art sequential solvers and therefore it is impossible to tell how well
do they scale for a large number of processors and current benchmarks.
The solver GridSAT~\cite{chrabakh2003gridsat} is run on a large heterogeneous grid of computers containing
hundreds of nodes
for several days and is reported to solve several (at that time) unsolved problems. Nevertheless, most
of those problems can now be solved by sequential solvers in a few minutes. Speedup results are not reported.
A recent grid-based solving method called Part-Tree-Learn~\cite{hyvarinen2011grid} 
is compared to Plingeling and is reported to solve less instances than Plingeling. 
This is despite the fact that in their comparison the number of processors 
available to Plingeling was slightly less~\cite{hyvarinen2011grid}.

To design a successful explicit partitioning parallel solver, complex load balancing issues must be solved.
Additionally, explicit partitioning clearly brings runtime and space overhead.
If the main motivation of explicit partitioning is to ensure that the search-spaces explored by the
solvers have no overlap, then we believe
that the extra work does not pay off and 
frequent clause sharing is enough to approximate the desired behavior
\footnote{According to our experiments only 2-6\% of the clauses are learned simultaneously by 
different solvers in a pure portfolio, which is an indication 
that the overlap of search-spaces is relatively small.}.
Moreover, in \cite{hyvarinen2012designing} the authors argue that plain partitioning approaches
can increase the expected runtime compared to pure portfolio systems. They prove that under reasonable
assumptions there is always a distribution that results in an increased expected runtime unless the
process of constructing partitions is ideal.


\section{Design Decisions}
\label{sec-dd}
In this section we provide an overview of the high level design decisions made
when designing our portfolio-based SAT solver HordeSat.
\myparagraph{Modular Design.} Rather than committing to any particular SAT solver we design an interface
that is universal and can be efficiently implemented by current state-of-the-art SAT solvers. This
results in a more general implementation and the possibility to easily add new SAT solvers to
our portfolio.
\myparagraph{Decentralization.} All the nodes in our parallel system are equivalent. There is no leader
or central node that manages the search or the communication. Decentralized design
allows more scalability and also simplifies the algorithm.
\myparagraph{Overlapping Search and Communication.} The search and the clause exchange procedures run in different
(hardware) threads in parallel. The system is implemented in a way that the search procedure never waits
for any shared resources at the expense of losing some of the shared clauses.
\myparagraph{Hierarchical Parallelization.} HordeSat is designed to run on clusters of computers (nodes) with
multiple processor cores, i.e., we have two levels of parallelization. 
The first level uses the shared memory model to
communicate between solvers running on the same node and the second level relies on message passing
between the nodes of a cluster.

The details and implementation of these points are discussed below.


\section{Black Box for Portfolios}
Our goal is to develop a general parallel portfolio solver based on existing
state-of-the-art sequential CDCL solvers without committing to any particular solver.
To achieve this we define a C++ interface that is used to access the solvers in the portfolio.
Therefore new SAT solvers can be easily added just by implementing this interface.
By \emph{core solver} we will mean a SAT solver implementing the interface.

In this section we describe the essential methods of the interface. All the methods
are required to be implemented in a thread safe way, i.e., safe execution by multiple 
threads at the same time must be guaranteed.
First we start with the basic methods which allow us to solve formulas and
interrupt the solver.

\myparagraph{\tt\bf {void addClause(vector$<$int$>$ clause)}:}
This method is used to load the initial formula that is to be solved.
The clauses are represented as lists of literals which are represented as integers in the usual way.
All the clauses must be considered by the solver at the next call of \texttt{solve}.
\myparagraph{\tt\bf SatResult solve():}
This method starts the search for the solution of the formula specified by the 
\texttt{addClause} calls. The return value is one
of the following \texttt{SatResult = \{SAT, UNSAT, UNKNOWN\}}.
The result \texttt{UNKNOWN} is returned when the solver is interrupted by 
calling \texttt{setSolverInterrupt()}.
\myparagraph{\tt\bf void setSolverInterrupt():}
Posts a request to the core solver instance to interrupt the search as soon as possible. If the method
\texttt{solve} has been called, it will return \texttt{UNKNOWN}. Subsequent calls
of \texttt{solve} on this instance must return \texttt{UNKNOWN} until the method \texttt{unsetSolverInterrupt} is called.
\myparagraph{\tt\bf void unsetSolverInterrupt():}
Removes the request to interrupt the search.

Using these four methods, a simple portfolio can be built. When using several
instances of the same deterministic SAT solver, some diversification can be
achieved by adding the clauses in a different order to each solver. 

More options for diversification are made possible via the following two
methods.  A good way of diversification is to set default \emph{phase} values for the
variables of the formula, i.e., truth values to be tried first. These are then
used by the core solver when selecting decision literals.  In general many
solver settings can be changed to achieve diversification. Since these may be
different for each core solver we define a general method for diversification
which the core solver can implement in its own specific way.

\myparagraph{\tt\bf void setPhase(int var, bool phase):}
This method is used to set a default phase of a variable. 
The solver is allowed to
ignore these suggestions.
\myparagraph{\tt\bf void diversify(int rank, int size):}
This method tells the core solver to diversify its settings. The specifics of diversification
are left to the solver. The provided parameters can be used by the solver to determine
how many solvers are working on this problem (\texttt{size}) and which one of those
is this solver (\texttt{rank}). A trivial implementation of this method could be to
set the pseudo-random number generator seed of the core solver to \texttt{rank}.

The final three methods of the interface deal with clause sharing. The solvers can produce and
accept clauses. Not all the learned clauses are shared. It is expected that 
each core solver initially offers only a limited
number of clauses which it considers most worthy of sharing. The solver should increase
the number of exported clauses when the method \texttt{increaseClauseProduction} is called.
This can be implemented by relaxing the constraints on the learned clauses selected
for exporting.

\myparagraph{\tt\bf void addLearnedClause(vector$<$int$>$ clause):}
This method is used to add learned clauses received from other solvers of the portfolio.
The core solver can decide when and whether the clauses added using this method are actually considered
during the search.
\myparagraph{\tt\bf void setLearnedClauseCallback(LCCallback* callback):}
This method is used to set a callback class that will process the clauses shared by this solver.
To export a clause, the core solver will call the \texttt{void write(vector<int> clause)}
method of the \texttt{LCCallback} class. Each clause exported by this
method must be a logical consequence of the clauses added using
\texttt{addClause} or \texttt{addLearnedClause}.
\myparagraph{\tt\bf void increaseClauseProduction():}
Inform the solver that more learned clauses should be shared. This could mean
for example that learned clauses of bigger size or higher glue value~\cite{audemard2009predicting}
will be shared. 


The interface is designed to closely match current CDCL SAT solvers, but any kind of SAT solver
can be used. For example a local search SAT solver could implement
the interface by ignoring the calls to the clause sharing methods.

For our experiments we implemented the interface by writing binding code for 
MiniSat \cite{sorensson2005minisat} and Lingeling \cite{biere13lingeling}. 
In the latter case no modifications to
the solver were required and the binding code only uses the incremental interface of Lingeling.
As for MiniSat, the code has been slightly modified to support the three clause sharing methods.


\section{The Portfolio Algorithm}
In this section we describe the main algorithm used in HordeSat.
As already mentioned in section~\ref{sec-dd} we use two levels of parallelization.
HordeSat can be viewed as a multithreaded program that communicates using
messages with other instances of the same program. The communication is implemented using
the Message Passing Interface (MPI) \cite{gropp1996mpi}. Each MPI process
runs the same multithreaded program 
and takes care about the following tasks:
\begin{itemize}
\item Start the core solvers using \texttt{solve}.
Use one fresh thread for each core solver.
\item Read the formula and add its clauses to each core solver using \texttt{addClause}.
\item Ensure diversification of the core solvers with respect to the other processes. 
\item Ensure that if one of the core solvers solves the problem all the other core solvers and
processes are notified and stopped. This is done by using \texttt{setSolverInterrupt} for each
core solver and sending a message to all the participating processes.
\item Collect the exported clauses from the core solvers, filter duplicates and send them to the other
processes. Accept the exported clauses of the other processes, filter them and distribute them 
to the core solvers. 
\end{itemize}

The tasks of reading the input formula, diversification, and solver starting are performed once after 
the start of the process. The communication of ending and clause exchange is performed 
periodically in rounds until a solution is found. The main thread sleeps between these rounds
for a given amount of time specified as a parameter of the solver (usually around 1 second).
The threads running the core solvers are working uninterrupted during the whole time of the search.

\subsection{Diversification}
Since we can only access the core solvers via the interface defined above, our only tools for diversification
are setting phases using the \texttt{setPhase} method and calling the solver specific
\texttt{diversify} method.

The \texttt{setPhase} method allows us to partition the search space in a semi-explicit fashion. An explicit
search space splitting into disjoint subspaces is usually done by imposing phase restrictions 
instead of just recommending them. The explicit approach is used in parallel solvers utilizing 
guiding paths \cite{chrabakh2003gridsat} and dynamic work stealing \cite{jurkowiak2005parallelization}.

We have implemented and tested the following diversification procedures based on literal phase
recommendations.
\begin{itemize}
\item \emph{Random}. Each variable gets a phase recommendation for each core solver randomly. 
Note that this is different
from selecting a random phase each time a decision is made for a variable in the CDCL procedure.
\item \emph{Sparse}. Each variable gets a random phase recommendation on exactly one of the host 
solvers in the entire portfolio. For the other solvers no phase recommendation is made for the given variable.
\item \emph{Sparse Random}. For each core solver each variable gets a random phase recommendation
with a probability of $(\#solvers)^{-1}$, where $\#solvers$ is the total number of core solvers in the portfolio.
\end{itemize}

Each of these can be used in conjunction with the \texttt{diversify} method whose behavior is defined by
the core solvers. As already mentioned we use Lingeling and MiniSat as core solvers. In case of MiniSat,
we implemented the \texttt{diversify} method by only setting the random seed. For Lingeling
we copied the diversification algorithm from Plingeling \cite{biere13lingeling}, 
which is the multi-threaded version of Lingeling based on the portfolio approach and the winner
of the parallel application track of the 2014 SAT Competition \cite{belov2014sat}.
In this algorithm 16 different parameters of Lingeling are used for diversification.

\subsection{Clause Sharing}
The clause sharing in our portfolio happens periodically in rounds. Each round a fixed sized
(1500 integers in the implementation)
message containing the literals of the shared clauses is exchanged by all the MPI processes in an 
all-to-all fashion.
This is implemented by using the \texttt{MPI\_Allgather} \cite{gropp1996mpi} 
collective communication routine defined by the MPI standard.

Each process prepares the message by collecting the learned clauses from its core solvers. The clauses
are filtered to remove duplicates. The fixed sized message buffer is filled
up with the clauses, shorter clauses are preferred. Clauses that did not fit are discarded.
If the buffer is not filled up to its full capacity 
then one of the core solvers of the process is requested to increase its clause production
by calling the \texttt{increaseClauseProduction} method.


The detection of duplicate clauses is implemented by using Bloom filters~\cite{bloom1970space}.
A~Bloom filter is a space-efficient probabilistic set data structure that allows false-positive matches, which
in our case means that some clauses might be considered to be duplicates even if they are not.
The usage of Bloom filters requires a set of hash functions that map clauses to integers. 
We use the following hash function which ensures that permuting the literals of a clause
does not change its hash value.
\begin{equation*}
H_i(C)=\bigoplus_{\ell\in C}\ell \cdot \Id{primes}[\Id{abs}(\ell\cdot i)\bmod |\Id{primes}|]
\end{equation*}
where $i>0$ is a parameter we are free to choose, $C$ is a clause, $\oplus$
denotes bitwise exclusive-or, and \Id{primes} is an array of large prime
numbers. Literals are interpreted as integers in the usual way, i.e., $x_j$ as
$j$ and $\overline{x}_j$ as $-j$.

Each MPI process maintains one Bloom filter $g_x$ for each of its core solvers $x$ and an additional global one $g$.
When a core solver $x$ exports a learned clause $C$, the following steps are taken.
\begin{itemize}
\item Clause $C$ is added to $g_x$.
\item If $C\not\in g$, $C$ is
added to $g$ as well as into a data structure $e$ for export.
\item If several core solvers concurrently try to access $e$, only one will
succeed and the new clauses of the other core solvers are ignored. 
This way, we avoid contention at the shared resource $e$ and rather ignore some clauses.
\end{itemize}

After the global exchange of learned clauses, the incoming clauses need to be
filtered for duplicates and distributed to the core solvers. The first task is
done by using the global Bloom filter $g$. For the second task we utilize the
thread local filters $g_x$ to ensure that each of them receives only new clauses.

All the Bloom filters are periodically reset, which allows the repeated sharing of clauses
after some time. Our initial experiments showed that this approach is more beneficial than
maintaining a strict ``no duplicate clauses allowed''-policy.

Overall, there are three reasons why a clause offered by a core solver can get discarded. 
One is that it was duplicate or wrongly
considered to be duplicate due to the probabilistic nature of Bloom filters. Second is that another
core solver was adding its clause to the data structure for global export at the same time.
The last reason is that it did not fit into the fixed size message sent to the other MPI processes.
Although important learned clauses might get lost, we believe that this relaxed approach is
still beneficial since it allows a simpler and more efficient implementation of clause sharing.

\section{Experimental Evaluation}
\label{sec-exp}
To examine our portfolio-based parallel SAT solver HordeSat we did experiments with two kinds of benchmarks.
We used the benchmark formulas from the application tracks of the 
2011 and 2014 SAT Competitions \cite{belov2014sat} (545 instances)
\footnote{Originally we only used the 2014 instances. A reviewer suggested to try the 2011 instances 
also, conjecturing that they would be harder to parallelize. Surprisingly, 
the opposite turned out to be true.}
and randomly generated 3-SAT formulas (200 sat and 200 unsat instances). 
The random formulas have 250--440 variables and 4.25 times as many clauses, which corresponds to
the phase transition of 3-SAT problems \cite{parkes97backbones}.

\begin{figure}[!t]
\centering
\includegraphics[width=0.8\linewidth]{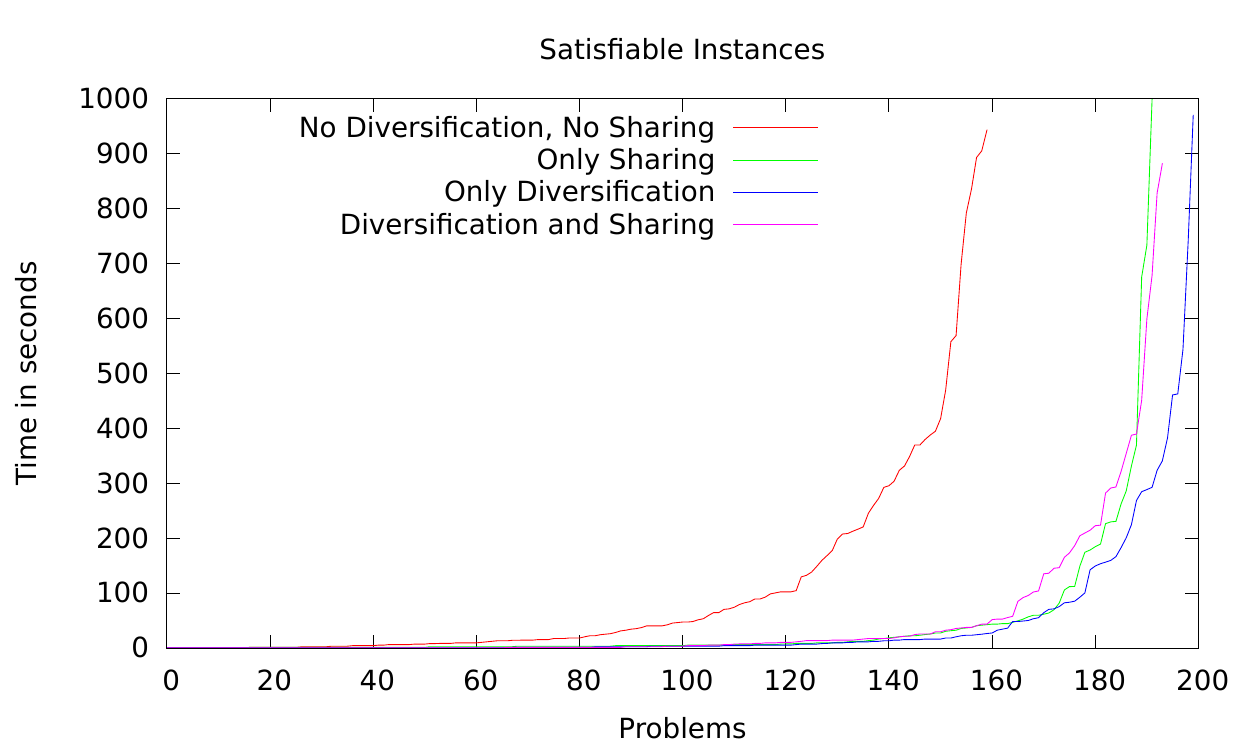}
\includegraphics[width=0.8\linewidth]{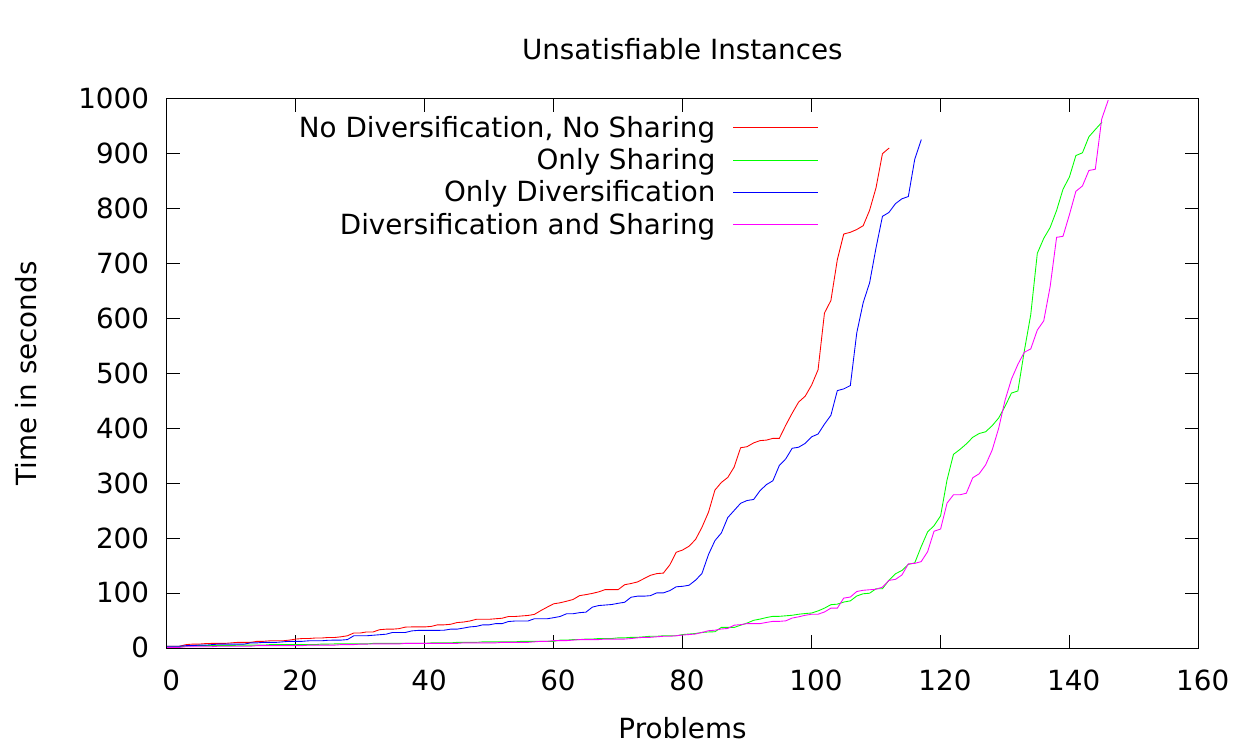}
\caption{The influence of diversification and clause sharing on the performance of HordeSat
using Lingeling (16 processes with 1 thread each) on random 3-SAT problems.}
\label{fig_rnd}
\end{figure}

The experiments were run on a cluster allowing us to reserve up to 
128 nodes. Each node has two octa-core Intel Xeon E5-2670 
processors (Sandy Bridge) with 2.6 GHz and 64 GB of main memory.
Therefore each node has 16 cores and the total number of available cores is 2048.
The nodes communicate using an InfiniBand 4X QDR Interconnect and use
the SUSE Linux Enterprise Server 11 (x86\_64) (patch level~3) operating system.
HordeSat was compiled using g++ (SUSE Linux) 4.3.4 [gcc-4\_3-branch revision 152973] with the ``-O3'' flag.

If not stated otherwise, we use the following parameters:
The time of sleeping between clause sharing rounds is 1 second.
The default diversification algorithm is the combination of
``sparse random'' and the native diversification of the core solver.
In the current version two core solvers are supported -- Lingeling and MiniSat.
The default value is Lingeling which is used in all the experiments presented below. It is also possible
to use a combination of Lingeling and MiniSat. Using only Lingeling gives by far the best results on
the used benchmarks.
The time limit per instance is 1\,000 seconds for parallel solvers 
and 50\,000 seconds for the sequential solver Lingeling.
Detailed results of all the presented experiments as well as the source code of
HordeSat and all the used benchmark problems can be found at \url{http://baldur.iti.kit.edu/hordesat}.

\subsection{Clause Sharing and Diversification}

We investigated the individual influence of clause sharing and diversification on the performance of
our portfolio. In the case of application benchmarks we obtained the unsurprising result that both
diversification and clause sharing are highly beneficial for satisfiable as well as unsatisfiable instances.
However, for random 3-SAT problems the results are more interesting.

By looking at the cactus plots in Figure \ref{fig_rnd} we can observe that clause sharing is essential for
unsatisfiable instances while not significant and even slightly detrimental for satisfiable problems.
On the other hand, diversification has only a small benefit for unsatisfiable instances.
This observation is related to a more general question of 
intensification vs diversification in parallel SAT solving \cite{DBLP:conf/cp/GuoHJS10}.

For the experiments presented in Figure \ref{fig_rnd} we used sparse diversification combined
with the \texttt{diversify} method, which in this case copies the behavior of Plingeling.
It is important to note that some diversification arises due to the non-deterministic nature
of Lingeling, even when we do not invoke it explicitly by using the \texttt{setPhase} or \texttt{diversify}
methods.

\subsection{Scaling on Application Benchmarks}
\label{sec-speedup}

\begin{figure}[t]
\centering
\includegraphics[width=0.8\linewidth]{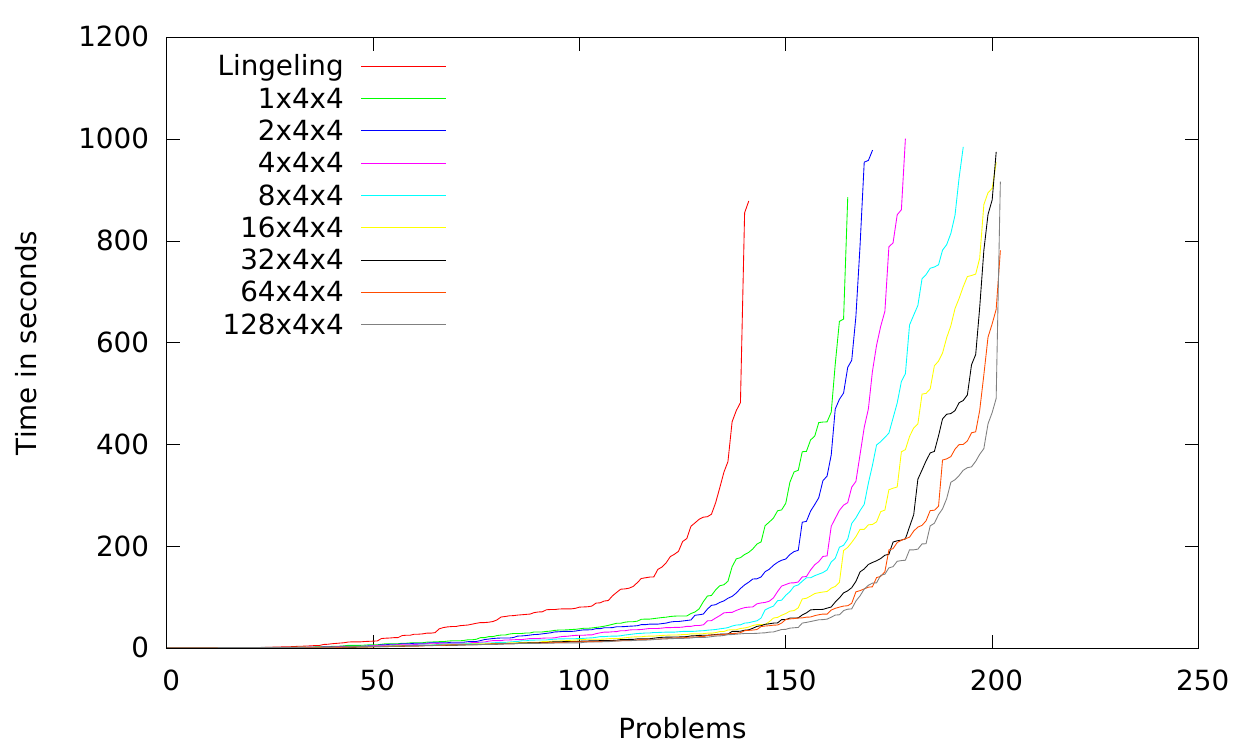}
\includegraphics[width=0.8\linewidth]{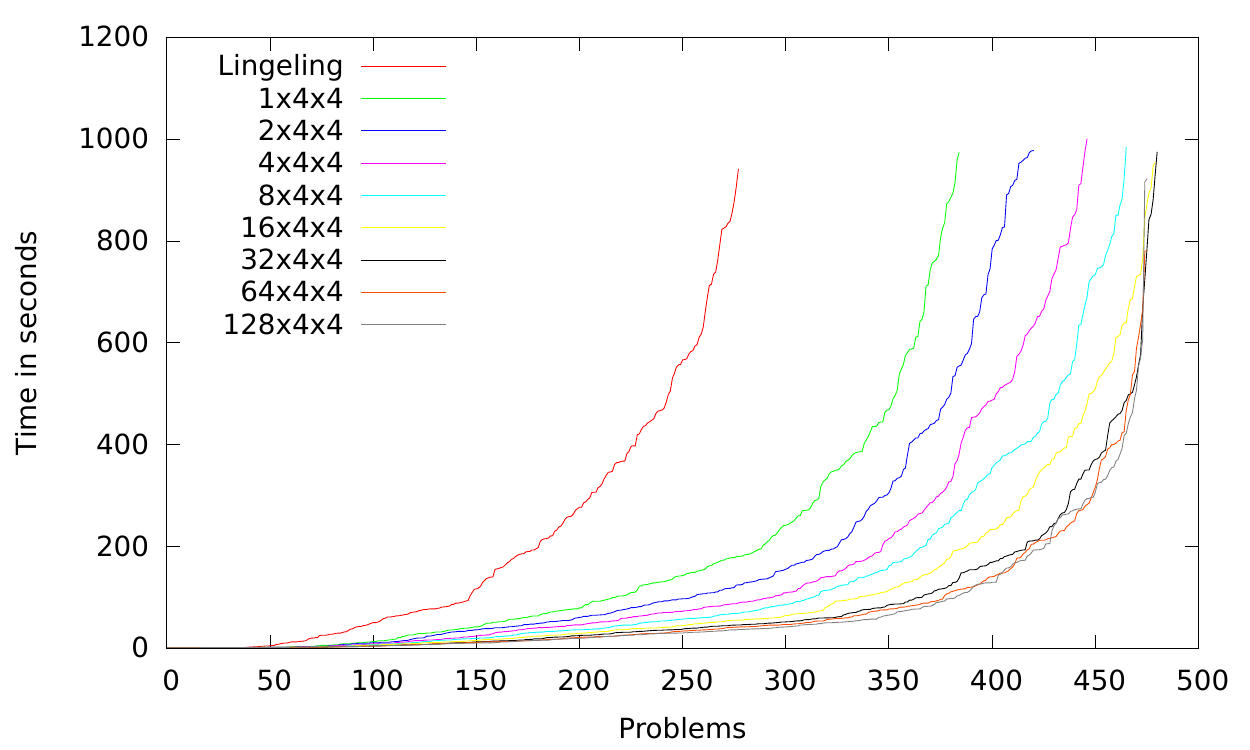}
\caption{The impact of doubling the number of processors on the runtime and
 the number solved problems for the 2011 and the union of 2011 and 2013 application instances. 
 The labels represent (\#nodes)x(\#processes/node)x(\#threads/process).}
\label{fig_app11_cact}
\end{figure}

\begin{table}[t]
\renewcommand{\arraystretch}{1.15}
\renewcommand\tabcolsep{4pt}
\begin{center}

\begin{tabular}{r || r | r | r  r  r |  r  r  r | r}
Core~~ & Parallel & Both~ & \multicolumn{3}{c|}{Speedup All} & \multicolumn{3}{c|}{Speedup Big}
& \\
Solvers & Solved~ & Solved & Avg. & Tot. & Med. & Avg. & Tot. & Med. & CBS \\
\hline
1x4x4 & 385 & 363 & 303 & 25.01 & 3.08 & 524 & 26.83 & 4.92 & 5.86 \\
2x4x4 & 421 & 392 & 310 & 30.38 & 4.35 & 609 & 33.71 & 9.55 & 22.44 \\
4x4x4 & 447 & 405 & 323 & 41.30 & 5.78 & 766 & 49.68 & 16.92 & 68.90 \\
8x4x4 & 466 & 420 & 317 & 50.48 & 7.81 & 801 & 60.38 & 32.55 & 102.27 \\
16x4x4 & 480 & 425 & 330 & 65.27 & 9.42 & 1006 & 85.23 & 63.75 & 134.37 \\
32x4x4 & 481 & 427 & 399 & 83.68 & 11.45 & 1763 & 167.13 & 162.22 & 209.07 \\
64x4x4 & 476 & 421 & 377 & 104.01 & 13.78 & 2138 & 295.76 & 540.89 & 230.37 \\
128x4x4 & 476 & 421 & 407 & 109.34 & 13.05 & 2607 & 352.16 & 867.00 & 216.69 \\
\hline
pling8 & 372 & 357 & 44 & 18.61 & 3.11 & 67 & 19.20 & 4.12 & 4.77 \\
pling16 & 400 & 377 & 347 & 24.83 & 3.53 & 586 & 26.18 & 5.89 & 7.34 \\
1x8x1 & 373 & 358 & 53 & 19.57 & 3.13 & 81 & 20.42 & 4.36 & 4.79 \\
1x16x1 & 400 & 376 & 325 & 27.78 & 4.06 & 548 & 30.30 & 6.98 & 7.34 \\
\end{tabular}

\end{center}
\caption{HordeSat configurations (\#nodes)x(\#processes/node)x(\#threads/process)
compared to Plingeling with a given number of threads.
The second column is the number of instances solved by the parallel solvers, the third is
the number of instances solved by both Lingeling and the parallel solver.
The following six columns contain the average, total, and median speedups for either all the instances
solved by the parallel solvers or only big instances (solved after 10(\#threads) seconds by Lingeling).
The last column contains the ``count based speedup'' values defined in Subsection~\ref{sec-speedup}.}
\label{tab-speedup}
\end{table}

\begin{figure}[t]
\centering
\includegraphics[width=0.8\linewidth]{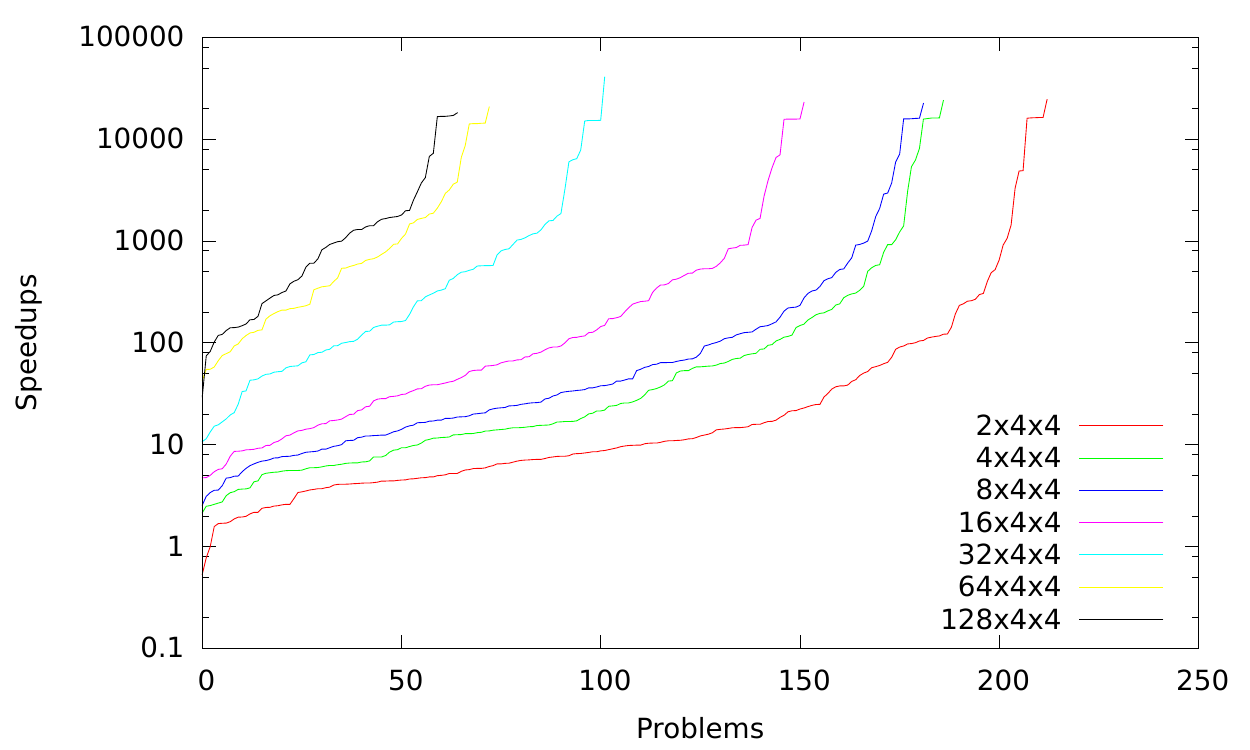}
\caption{Distribution of speedups on the ``big instances'' -- the data corresponding to
Columns~7--9 of Table~\ref{tab-speedup}.}
\label{fig_app_speedupcact}
\end{figure}

In parallel processing, one usually wants good scalability in the sense that the
speedup over the best sequential algorithm goes up near linearly with the number
of processors. Measuring scalability in a reliable and meaningful way is
difficult for SAT solving since running times are highly nondeterministic.
Hence, we need careful experiments on a large benchmark set chosen in an
unbiased way. We therefore use the application benchmarks of the 2011 and 2014 Sat
Competitions. Our sequential reference is Lingeling which won the most recent (2014) competition.  
We ran experiments using 1,2,4,\ldots,512 processes with four threads each, each cluster nodes
runs 4 processes. The results are summarized in Figure~\ref{fig_app11_cact} using cactus plots.
We can observe that increased parallelism is always beneficial for the 2011 benchmarks. In the
case of all the benchmarks the benefits beyond 32 nodes are not apparent.

From a cactus plot it is not easy to see whether the additional performance is a
reasonable return on the invested hardware resources. Therefore Table~\ref{tab-speedup}
summarizes that information in several ways in order to quantify the overall
scalability of HordeSat on the union of the 2011 and 2013 benchmarks. 
We compute speedups for all the instances solved by the parallel solver.
For instances not solved by Lingeling within its time limit $T=50\,000s$
we generously assume that it would solve them
if given $T+\epsilon$ seconds and use the runtime of $T$ for speedup calculation.
Column~4 gives the average of these values. We observe considerable
superlinear speedups \emph{on average} for all the configurations tried.  However, this
average is not a very robust measure since it is highly dependent on a few very
large speedups that might be just luck. In Column~5 we show the total speedup, which is the
sum of sequential runtimes divided by the sum of parallel runtimes and Column~6 contains the median speedup.

Nevertheless, these figures treat HordeSat unfairly since most instances are actually too easy for
investing a lot of hardware. Indeed, in parallel computing, it is usual to
analyze the performance on many processors using \emph{weak scaling} where one
increases the amount of work involved in the considered instances proportionally
to the number of processors. Therefore in columns 7--9 we
restrict ourselves to those instances where Lingeling needs at least 10$p$ seconds where $p$ is the
number of core solvers used by HordeSat. The average speedup gets considerably
larger as well as the total speedup, especially for the large configurations.
The median speedup also increases but remains slightly sublinear.
Figure~\ref{fig_app_speedupcact} shows the distribution of speedups for these instances.

Another way to measure speedup robustly is to compare
the times needed to solve a given number of instances.  Let $T_1$ ($T_p$) denote
the per instance time limits of the sequential (parallel) solver (50\,000s
(1\,000s) in our case).  Let $n_1$ ($n_p$) denote the number of instances solved
by the sequential (parallel) solver within time $T_1$ ($T_p$). If $n_1\geq n_p$
($n_1< n_p$) let $T_1'$ ($T_p'$) denote the smallest time
limit for the sequential (parallel) solver such that it solves $n_p$ ($n_1$)
instances within the time limit $T_1'$ ($T_p'$).
We define the \emph{count based speedup} (CBS) as 
$$\text{CBS}=\begin{cases}
T_1/T_p' & \text{ if }   n_1<n_p\\
T_1'/T_p & \text{ otherwise }.
\end{cases}$$

The CBS scales almost linearly up to 512 cores and stagnates afterward. 
We are not sure whether this indicates a scalability limit of HordeSat
or rather reflects a lack of sufficiently difficult instances -- in our collection, 
there are only 65 eligible instances.

\subsection{Comparison with Plingeling}

\begin{figure}[!t]
\centering
\includegraphics[width=0.8\linewidth]{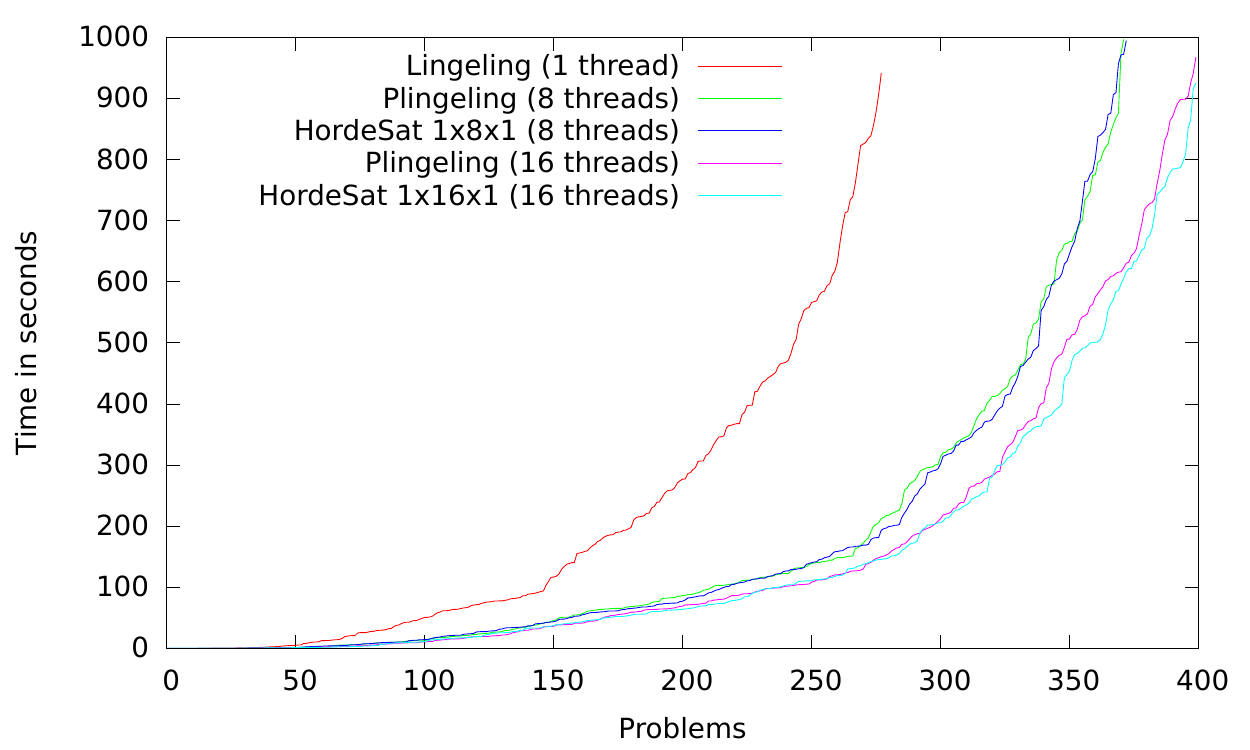}
\caption{Comparison of HordeSat and Plingeling with Lingeling on the 2011 and 2014 Sat Competition
benchmarks.}
\label{fig_pling}
\end{figure}

The most similar parallel SAT solver to our portfolio is the state-of-the-art solver Plingeling \cite{biere13lingeling}.
Plingeling is the winner of the parallel track of the 2014 SAT Competition.
Both solvers are portfolio-based, they are using Lingeling and even some diversification code is shared.
The main differences are in the clause sharing algorithms and that Plingeling does not run on
clusters only single computers. For this reason we can compare the two solvers only on a single
node. The results of this comparison on the benchmark problems of the 2011 and 2014 SAT Competition
are displayed in Figure~\ref{fig_pling}. Speedup values are given in Table~\ref{tab-speedup}.

Both solvers significantly outperform Lingeling. The performance of HordeSat and Plingeling is almost
indistinguishable when running with 8 cores, while on 16 cores HordeSat gets slightly ahead of Plingeling.

\section{Conclusion}

HordeSat has the potential to reduce solution times of difficult yet solvable
SAT instances from hours to minutes using hundreds of cores on commodity
clusters. This may open up new interactive applications of SAT solving.  We find
it surprising that this was achieved using a relatively simple, portfolio based
approach that is independent of the underlying core solver. In particular, this
makes it likely that HordeSat can track future progress of sequential SAT
solvers.


The Sat solver that works best with HordeSat for application benchmarks is Lingeling. Plingeling is another
parallel portfolio solver based on Lingeling and it is also the winner of the most recent (2014) Sat Competition.
Comparing the performance of HordeSat and Plingeling reveals that HordeSat is almost indistinguishable when
running with 8 cores and slightly outperforms Plingeling when running with 16 cores. This demonstrates that
there is still room for the improvement of shared memory based parallel portfolio solvers.

Our experiments on a cluster with up to 2048 processor cores 
show that HordeSat is scalable in highly parallel environments. We observed superlinear and nearly linear
scaling in several measures such as average, total, and median speedups, particularly on hard instances.
In each case increasing the number of available cores resulted in significantly reduced runtimes. 

\subsection{Future Work}
An important next step is to work on the scalability of HordeSat for 1024
cores and beyond.  This will certainly involve more adaptive clause
exchange strategies. Even for single node configurations, low level
performance improvements when using modern machines with dozens of
cores seem possible.  We also would like to investigate what benefits
can be gained by having a tighter integration of core solvers by
extending the interface.  Including other kinds of (not necessarily
CDCL -- based) core solvers might also bring improvements.

When considering massively parallel SAT solving we probably have to move to even
more difficult instances to make that meaningful. When this also means
\emph{larger} instances, memory consumption may be an issue when running many
instances of a SAT solver on a many-core machine. Here it might be interesting
to explore opportunities for sharing data structures for multiple SAT solvers or
to decompose problems into smaller subproblems by recognizing their structure.

\subsubsection*{Acknowledgment.} 
We would like to thank Armin Biere for fruitful discussion about
the usage of the Lingeling API in a parallel setting.


\newpage

\bibliographystyle{splncs03}

\bibliography{references}

\begin{thebibliography}{10}
\providecommand{\url}[1]{\texttt{#1}}
\providecommand{\urlprefix}{URL }

\bibitem{audemard2009predicting}
Audemard, G., Simon, L.: Predicting learnt clauses quality in modern sat
  solvers. In: IJCAI. vol.~9, pp. 399--404 (2009)

\bibitem{audemard2014lazy}
Audemard, G., Simon, L.: Lazy clause exchange policy for parallel sat solvers.
  In: Theory and Applications of Satisfiability Testing--SAT 2014, pp.
  197--205. Springer (2014)

\bibitem{belov2014sat}
Belov, A., Diepold, D., Heule, M.J., J{\"a}rvisalo, M.: Sat competition 2014
  (2014)

\bibitem{biere13lingeling}
Biere, A.: Lingeling, plingeling and treengeling entering the sat competition
  2013. In: In Proceedings of SAT Competition 2013, A. Balint, A. Belov, M. J.
  H. Heule, M. J\"arvisalo (editors), vol. B-2013-1 of Department of Computer
  Science Series of Publications B pages 51-52, University of Helsinki, 2013.
  (2013)

\bibitem{biere2009conflict}
Biere, A., Heule, M., van Maaren, H., Walsh, T.: Conflict-driven clause
  learning sat solvers. Handbook of Satisfiability, Frontiers in Artificial
  Intelligence and Applications pp. 131--153 (2009)

\bibitem{blochinger2005zetasat}
Blochinger, W., Westje, W., Kuchlin, W., Wedeniwski, S.: Zetasat-boolean
  satisfiability solving on desktop grids. In: Cluster Computing and the Grid,
  2005. CCGrid 2005. IEEE International Symposium on. vol.~2, pp. 1079--1086.
  IEEE (2005)

\bibitem{bloom1970space}
Bloom, B.H.: Space/time trade-offs in hash coding with allowable errors.
  Communications of the ACM  13(7),  422--426 (1970)

\bibitem{chrabakh2003gradsat}
Chrabakh, W., Wolski, R.: Gradsat: A parallel sat solver for the grid. In:
  Proceedings of IEEE SC03 (2003)

\bibitem{chrabakh2003gridsat}
Chrabakh, W., Wolski, R.: Gridsat: A chaff-based distributed sat solver for the
  grid. In: Proceedings of the 2003 ACM/IEEE conference on Supercomputing.
  p.~37. ACM (2003)

\bibitem{flanagan2003theorem}
Flanagan, C., Joshi, R., Ou, X., Saxe, J.B.: Theorem proving using lazy proof
  explication. In: Computer Aided Verification. pp. 355--367. Springer (2003)

\bibitem{gil2008pmsat}
Gil, L., Flores, P., Silveira, L.M.: Pmsat: a parallel version of minisat.
  Journal on Satisfiability, Boolean Modeling and Computation  6,  71--98
  (2008)

\bibitem{gropp1996mpi}
Gropp, W., Lusk, E., Doss, N., Skjellum, A.: A high-performance, portable
  implementation of the mpi message passing interface standard. Parallel
  computing  22(6),  789--828 (1996)

\bibitem{DBLP:conf/cp/GuoHJS10}
Guo, L., Hamadi, Y., Jabbour, S., Sais, L.: Diversification and intensification
  in parallel {SAT} solving. In: Cohen, D. (ed.) Principles and Practice of
  Constraint Programming - {CP} 2010. Lecture Notes in Computer Science, vol.
  6308, pp. 252--265. Springer (2010)

\bibitem{hamadi2008manysat}
Hamadi, Y., Jabbour, S., Sais, L.: Manysat: a parallel sat solver. Journal on
  Satisfiability, Boolean Modeling and Computation  6,  245--262 (2008)

\bibitem{holldobler2011short}
H{\"o}lldobler, S., Manthey, N., Nguyen, V., Stecklina, J., Steinke, P.: A
  short overview on modern parallel sat-solvers. In: Proceedings of the
  International Conference on Advanced Computer Science and Information
  Systems. pp. 201--206 (2011)

\bibitem{hyvarinen2011grid}
Hyv{\"a}rinen, A.E., Junttila, T., Niemel{\"a}, I.: Grid-based sat solving with
  iterative partitioning and clause learning. In: Principles and Practice of
  Constraint Programming--CP 2011, pp. 385--399. Springer (2011)

\bibitem{hyvarinen2014incorporating}
Hyv{\"a}rinen, A.E., Junttila, T., Niemela, I.: Incorporating clause learning
  in grid-based randomized sat solving. Journal on Satisfiability, Boolean
  Modeling and Computation  6,  223--244 (2014)

\bibitem{hyvarinen2012designing}
Hyv{\"a}rinen, A.E., Manthey, N.: Designing scalable parallel sat solvers. In:
  Theory and Applications of Satisfiability Testing--SAT 2012, pp. 214--227.
  Springer (2012)

\bibitem{inprocessing}
J{\"a}rvisalo, M., Heule, M.J.H., Biere, A.: Inprocessing rules. In:
  Proceedings of IJCAR 2012. LNCS, vol. 7364, pp. 355--370. Springer (2012)

\bibitem{jurkowiak2005parallelization}
Jurkowiak, B., Li, C.M., Utard, G.: A parallelization scheme based on work
  stealing for a class of sat solvers. Journal of Automated Reasoning  34(1),
  73--101 (2005)

\bibitem{kautz1992planning}
Kautz, H.A., Selman, B., et~al.: Planning as satisfiability. In: ECAI. vol.~92,
  pp. 359--363 (1992)

\bibitem{aigs}
Kuehlmann, A., Paruthi, V., Krohm, F., Ganai, M.K.: Robust boolean reasoning
  for equivalence checking and functional property verification. IEEE
  Transactions on Computer-Aided Design of Integrated Circuits and Systems
  21(12) (2002)

\bibitem{marques1999grasp}
Marques-Silva, J.P., Sakallah, K.A.: Grasp: A search algorithm for
  propositional satisfiability. Computers, IEEE Transactions on  48(5),
  506--521 (1999)

\bibitem{martins2012overview}
Martins, R., Manquinho, V., Lynce, I.: An overview of parallel sat solving.
  Constraints  17(3),  304--347 (2012)

\bibitem{moskewicz2001chaff}
Moskewicz, M.W., Madigan, C.F., Zhao, Y., Zhang, L., Malik, S.: Chaff:
  Engineering an efficient sat solver. In: Proceedings of the 38th annual
  Design Automation Conference. pp. 530--535. ACM (2001)

\bibitem{ohmura2009csat}
Ohmura, K., Ueda, K.: c-sat: A parallel sat solver for clusters. In: Theory and
  Applications of Satisfiability Testing-SAT 2009, pp. 524--537. Springer
  (2009)

\bibitem{parkes97backbones}
Parkes, A.J.: Clustering at the phase transition. In: In Proc. of the 14th Nat.
  Conf. on AI. pp. 340--345. AAAI Press / The MIT Press (1997)

\bibitem{schulz2010parallel}
Schulz, S., Blochinger, W.: Parallel sat solving on peer-to-peer desktop grids.
  Journal of Grid Computing  8(3),  443--471 (2010)

\bibitem{sorensson2005minisat}
Sorensson, N., Een, N.: Minisat v1.13 a sat solver with conflict-clause
  minimization. SAT  2005 (2005)

\bibitem{hydra}
Xu, L., Hoos, H., Leyton-Brown, K.: Hydra: Automatically configuring algorithms
  for portfolio-based selection. AAAI Conference on Artificial Intelligence
  (2010)

\end{thebibliography}


\end{document}